\documentclass[11pt,aps,pra,superscriptaddress,showpacs,floatfix]{revtex4}
\usepackage{psfrag}
\usepackage{amsmath,amssymb}
\usepackage{type1cm}
\usepackage{bm}
\usepackage{comment}
\usepackage{mathptmx}
\usepackage{wrapfig}
\usepackage{multirow}
\usepackage[dvipdfmx]{graphicx}

\newcommand{\tlB}{\mbox{${\tilde B}$}}
\newcommand{\tlC}{\mbox{${\tilde C}$}}

\newcommand{\vx}{\bm{x}}

\newcommand{\br}{{\bf r}}

\newcommand{\vbr}{\mbox{$\bm{r}$}}

\newcommand{\vu}{\mbox{$\bm{u}$}}

\newcommand{\vlambda}{\mbox{$\bm{\lambda}$}}
\newcommand{\tvlamb}{\mbox{$^t{\bm{\lambda}}$}}
\newcommand{\dvlamb}{\mbox{$\bm{\dot \lambda}$}}
\newcommand{\tdvlamb}{\mbox{$^t{\bm{\dot \lambda}}$}}

\begin{document}

\title{Deformation Dependence of Breathing Oscillations 
\\in Bose - Fermi Mixtures at Zero Temperature}

\author{Tomoyuki~Maruyama}
\affiliation{College of Bioresource Sciences,
Nihon University,
Fujisawa 252-0880, Japan}
\affiliation{Advanced Science Research Center,
Japan Atomic Energy Agency, Tokai 319-1195, Japan}
\affiliation{National Astronomical Observatory of Japan, 
2-21-1 Osawa, Mitaka, Tokyo 181-8588, Japan}

\author{Tike's Yamamoto}
\affiliation{
Department of Physics,
Ritsumeikan University, Kusatsu 525-8577, Japan}

\author{Takushi Nishimura}
\affiliation{Department of Physics, Tokyo Metropolitan University, 
 Hachioji, Tokyo 192-0397, Japan}

\author{Hiroyuki Yabu}
\affiliation{
Department of Physics,
Ritsumeikan University, Kusatsu 525-8577, Japan}

\date{\today}

\begin{abstract}
We study the breathing oscillations in bose-fermi mixtures
 in the axially-symmetric deformed trap
of prolate, spherical and oblate shapes, 
and clarify the deformation dependence of the frequencies 
and the characteristics of collective oscillations. 
The collective oscillations of the mixtures in deformed traps 
are calculated in the scaling method. 
In largely-deformed prolate and oblate limits 
and spherical limit, 
we obtain the analytical expressions of the collective frequencies.
The full calculation shows that the collective oscillations become consistent with the analytically-obtained frequencies 
when the system is deformed into both prolate and oblate regions.
The complicated changes of oscillation characters are shown to occur 
in the transcendental regions around the spherically-deformed region.
We find that these critical changes of oscillation characters are explained by the level crossing behaviors 
of the intrinsic oscillation modes. 
The approximate expressions are obtained for the level crossing points 
that determine the transcendental regions.
We also compare the results of the scaling methods 
with those of the dynamical approach.
\end{abstract}


\maketitle

\section{Introduction}
\label{intr}

Study of the ultracold atomic gases have been made many progresses 
in many-body quantum theory 
through the Bose-Einstein condensates (BEC) \cite{nobel,Dalfovo,becth,Andersen}, 
two boson mixtures  \cite{Truscott,B-B}, 
degenerate atomic Fermi gases \cite{ferG},
and Bose-Fermi (BF) mixing gases \cite{Truscott,Schreck,BferM,Modugno}.
In particular, 
the BF mixtures, 
a typical example in which particles obeying different quantum statistics are intermingled,
attract physical interest 
to obtain new knowledge of many-body quantum systems 
because of a large variety of combinations of atomic species 
and controllability of the atomic interactions 
using the Feshbach resonance \cite{Fesh}.

A variety of theoretical studies have been done on the BF mixtures:
static properties \cite{Molmer,Amoruso,MOSY,Bijlsma,Vichi,Vichi1}, 
phase diagrams and phase separation phenomena \cite{Nygaard,Yi,Viverit,Capuzzi02}, 
instability  \cite{MSY1,Roth,Capuzzi03} and
collective excitations \cite{MSY,Minguzzi,MOSY,zeros,Yip,sogo,tomoBF,monoEX,tbDPL,QPBF,MY12}. 

The researches on the trapped atomic gases of the Yb isotopes
have been performed by Kyoto university group:
the BEC \cite{Takasu} and the Fermi-degeneracy \cite{Fukuhara1}.  
They also succeeded in realizing
the BF mixtures of the isotopes. 
The Yb consists of many kinds of isotopes--- 
five bosons (${}^{168,170,172,174,176}$Yb) 
and two fermions (${}^{171,173}$Yb), 
which give a variety of combinations in the BF mixtures.
The scattering lengths of the boson-fermion interactions have been determined
experimentally by the group \cite{Fukuhara1,Fukuhara2,Kitagawa,Fukuhara3,Fukuhara-M}, 
and the experimental studies of the ground state properties and the collective oscillations 
of the BF mixtures is now under progressing.

In many characteristic properties, 
the spectrum of the collective excitations is
an important diagnostic signal of these systems;
they commonly appear in many-particle systems and 
are often sensitive to the inter-particle interaction and 
the structure of the ground and excited states.  
Using a dynamical approach to solve time-developments 
of the oscillation states 
with the time-dependent Gross-Pitaevskii (TDGP) and Vlasov equations, 
we have studied the monopole \cite{tomoBF,monoEX}, 
dipole \cite{tbDPL} and quadrupole \cite{QPBF} oscillations of the BF mixtures
in the spherical traps and the breathing oscillations 
in the largely prolate deformed system \cite{MY12}.
In these works, we found that 
the oscillational motions of the BF mixtures include various modes, 
and that, though the intrinsic frequencies are somewhat different from those 
in the sum-rule \cite{MOSY,MSY},
the dynamical calculations give consistent results 
with the random phase approximation (RPA) \cite{zeros,sogo} 
in the early stage of time development.

The oscillations are decoupled into the longitudinal and transverse  
oscillation modes in the largely prolate deformed system,
and into the monopole and quadrupole oscillation modes
in the completely spherical system.
However, the mixing behavior of these modes is not known 
in the deformed region between the above two extreme conditions.

We studied the deformation dependence 
of breathing oscillations in the two-component fermi gases
using the scaling approach\cite{scalTF2},
and found that 
the breathing oscillations can be described as superpositions of 
the four oscillation modes: the longitudinal and transverse oscillations 
with in-phase and out-of-phase\cite{scalTF}.
In the aspect of the mixing of these modes, 
the characteristic mixing behaviors for the breathing oscillations 
are found in the oblate, spherical and prolate traps, respectively. 

In BF mixtures, it should be noted that the in-phase and out-of-phase oscillations are not 
intrinsic modes in oscillations because the two components 
obey different statistics.
Thus, the BF mixtures are expected to show different behaviors 
in the mode mixing and resonances at the level crossing points from 
the two-component fermi gases.

In this work, then, we investigate the deformation dependence 
in the breathing oscillations of the BF mixtures by
examining  mixing properties of the longitudinal and
transverse oscillations.
For this purpose we use  the scaling approach, 
which is an approximation of the RPA \cite{scal,scalDP,bohi},
and describe  collective oscillations with macroscopic picture.
This method have reproduced the intrinsic frequencies of the BF
breathing oscillations  calculated in the 
time-dependent simulation \cite{MY12},
and can clearly demonstrate the mixing properties,

Though this scaling method is thought to describe 
the collective oscillations within the linear response regime,
the nonlinear calculation was applied 
to the oscillations of BEC \cite{CD96}, 
which reproduced the experimental results by Ref.~\cite{Mewes96} very well. 
Also, the method was used in the study of the oscillations of 
non-condensed bose gases \cite{KSS96} and the quasi-low dimension gases
\cite{QLDF}. 
In the next section, 
we give a brief explanation of the scaling method  
and discuss the mixing of the boson and fermion oscillations
in extreme conditions.
In Sec. III, 
we show the results in various deformed traps in the scaling method.
In Sec. IV,
we compare the collective oscillation modes obtained in the scaling method< 
with those in the dynamical approach based on the TDGP+Vlasov equation. 
The summary of the present work is given in Sec. V.

\newpage

\section{Collective Oscillations in the Scaling Method}
\label{Te}

Let's consider the oscillation behaviors of the mixture 
in the deformed system, and take up the collective modes  
from the macroscopic point of view.

 In order to calculate the oscillation modes, 
we use the scaling method, 
for which we only give a brief explanation, 
so that, for the details,  
the readers should consult the previous paper.~\cite{MY12}. 
Based on the method, 
we also discuss the mixing behaviors of the boson and fermion oscillations, 
which has not been done in detail before.

\subsection{Total Energy of BF mixtures and Scaled Variables}

Let's consider the zero-temperature BF mixture of dilute boson and 
one-component-fermion gases trapped  
in an axially symmetric potential of the symmetry $z$-axis;
the interatomic interactions are assumed to be zero-range 
with no fermion-fermion interactions. 
In the Hartree-Fock approximation, 
the total energy is represented 
by a functional of the condensed boson wave function $\phi_c$, 
and the $n$-th fermion single-particle wave functions $\psi_n$:
\begin{eqnarray}
     E_T &=&\int d^3{r} \Bigg[
            \frac{\hbar^2}{2M_B} N_b \nabla_r \phi_c^\dagger(\vbr) 
                         \nabla_r \phi_c(\vbr)
            +\frac{M_B}{2} (\Omega_T^2 \vbr_T^2 +\Omega_L^2 z^2)
                         N_b \phi_c^\dagger(\vbr) \phi_c(\vbr) 
%
%
            + \frac{2\pi \hbar^2 a_{BB}}{M_{Bu}} \{ N_b \phi_c^\dagger(\vbr) 
                                 \phi_c (\vbr)      \}^2  
\nonumber \\
         &&\qquad\quad
         +\frac{\hbar^2}{2M_f} \sum_{n=1}^{N_f} \nabla_r \psi_n^\dagger(\vbr) 
                                 \nabla_r \psi_n (\vbr)
         +\frac{1}{2} M_F \omega_f^2  (\Omega_T^2 \vbr_T^2 +\Omega_L^2 z^2)
          \sum_{n=1}^{N_f} \psi_n^\dagger(\vbr) \psi_n (\vbr) 
\nonumber \\
         &&\qquad\quad
         +2 \pi \hbar^2 \frac{M_B + M_F}{M_B M_F} a_{BF} 
         N_b \phi_c^\dagger (\vbr) \phi_c (\vbr) 
          \sum_{n=1}^{N_f} \psi_n^\dagger (\vbr) \psi_n (\vbr) \Bigg] ,
\label{etot}
\end{eqnarray}
where $M_{B,F}$ are the boson and fermion masses,
$a_{BB,BF}$ are the s-wave scattering lengths 
of the boson-boson and boson-fermion interactions,
and $N_{b,f}$ are the number of bosons and fermions. 
We assume the trapping potentials of the similar harmonic shape
with the transverse and longitudinal trapping frequencies $(\Omega_T, \Omega_L)$ for bosons 
and $(\omega_f \Omega_T, \omega_f \Omega_L)$ for fermions.
For future convenience, 
we define the frequency parameters:
$\Omega_B \equiv (\Omega_T^2\Omega_L)^{1/3}$, 
$\omega_T = \Omega_T/\Omega_B$, and
 $\omega_L = \Omega_L/\Omega_B = \omega_T^{-1/2}$.

Here, we use the time coordinate $\tau$ scaled with $\Omega_B^{-1}$.
The scaled wave functions $\phi_{\lambda}(\vbr, t)$ and $\psi_{\lambda,n}(\vbr,\tau)$ for bosons and fermions 
are introduced from the ground-state one-body wave functions, 
$\phi_c^{(g)}$ and $\psi^{(g)}_n$:
\begin{eqnarray}
     \phi_{\lambda}(\vbr, t) &=& 
          e^{i M_B \xi_B (\br,t) /\hbar} 
          e^{\lambda_{BT}(t) +\frac{1}{2}\lambda_{BL}(t)} 
          \phi_c^{(g)}(e^{\lambda_{BT}(t)} \vbr_T; e^{\lambda_{BL}(t)} z),
\label{scwfB}\\
     \psi_{\lambda,n}(\vbr,\tau) &=& 
          e^{i M_F \xi_F (\br,t)/\hbar} 
          e^{\lambda_{FT}(t) + \frac{1}{2}\lambda_{FL}(t)} 
          \psi_n^{(g)}(e^{\lambda_{FT}(t)} \vbr_T; e^{\lambda_{FL}(t)} z)
\label{scwfF}
\end{eqnarray}
with
\begin{equation}
     \xi_a(\vbr,\tau) = \frac{1}{2} 
          \left\{ {\dot \lambda}_{Ta}(\tau) \vbr_T^2  
                +{\dot \lambda}_{La}(\tau) z^2 \right\}, \qquad
     (a=B,F)
\end{equation}
where the collective coordinates,
$\lambda_{BT}$, $\lambda_{BL}$, $\lambda_{FT}$, $\lambda_{FL}$, 
boson transverse breathing (BTB), 
describes the boson longitudinal breathing (BLB), 
fermion transverse breathing (FTB)
and fermion longitudinal breathing (FLB) oscillation modes respectively,
and ${\dot \lambda}$'s are the time-derivatives.
It turns out later that these four modes are completely decoupled
in the BF mixtures of no interactions: $a_{BB} = a_{BF} =0$.

Using the wave-functions (\ref{scwfB}) and (\ref{scwfF})
for the total energy functional (\ref{etot}),
we obtain,
\begin{eqnarray}
     E_T &=& \frac{M_B}{2 \hbar^2}  \int d^3 r\, (\nabla_r \xi_B)^2 \rho_B 
          +  \frac{1}{2} e^{2 \lambda_{BT}} ( T_{B,1} + T_{B,2} )
          +  \frac{1}{2} e^{2 \lambda_{BL}}  T_{B,3} 
\nonumber \\
         &+& \frac{M_B}{2} \Omega_B^2 \int d^3 r 
                \left\{ \omega_T^2 e^{- 2 \lambda_{BT} } (r_1^2 + r_2^2)
                       +\omega_L^2 e^{-2 \lambda_{BL}} z^2 \right\} 
                \rho_B (\vbr) 
\nonumber \\
         &+& \frac{2\pi \hbar^2 a_{BB}}{M_{BB}} 
             e^{2 \lambda_{BT} + \lambda_{BL} }
             \int d^3 r\,  \rho_B^2 (\vbr) , 
\nonumber \\
           &+& \frac{M_F}{2 \hbar^2} \int d^3 r\, (\nabla_r \xi_F)^2 \rho_F 
            +  \frac{1}{2} e^{2 \lambda_{FT}} ( T_{F,1} + T_{F,2} )
            +  \frac{1}{2} e^{2 \lambda_{FL}}  T_{F,3}
\nonumber \\
           &+& \frac{M_f}{2} \omega_f^2 \Omega_B^2 \int d^3 r 
               \left\{ \omega_T^2 e^{- 2 \lambda_{FT} } (r_1^2 + r_2^2)
            +  \omega_L^2e^{-2 \lambda_{FT}} z^2 ) \right\} 
               \rho_F (\vbr) , 
\nonumber \\
          &+& 2 \pi \hbar^2 \frac{M_B + M_F}{M_B M_F} a_{BF} 
              e^{2 \lambda_{BT} +\lambda_{BL} 
                +2 \lambda_{FT} +\lambda_{FL}} 
              \int d^3 r\, 
              \rho_B(e^{\lambda_{BT}} \vbr_T; e^{\lambda_{BL}} z)
              \rho_F(e^{\lambda_{FT}} \vbr_T; e^{\lambda_{FL}} z),
\label{etotTFs}
\end{eqnarray}
where
\begin{eqnarray*}
     T_{B,i} &=& \frac{\hbar^2}{M_B}
                 \int d^3 r  \left| \frac{\partial}{\partial r_i} \phi (\vbr) \right|^2,
     \qquad (i=1,2,3)
\\
     T_{F,i} &=&  \frac{\hbar^2}{M_F} \sum_n \int d^3{r}
       \left| \frac{\partial}{\partial r_i} \psi_n (\vbr) \right|^2, 
     \qquad (i=1,2,3)
\\
     \rho_B (\vbr) &=& N_b |\phi_c (\vbr)|^2, \quad
     \rho_F (\vbr)  = \sum^{N_f}_{n=1} |\psi_n (\vbr)|^2.
\end{eqnarray*} 

In order to obtain the ground state of the system,
we use the Thomas-Fermi (TF) approximation:
\begin{equation}
     T_B (\vbr) =0,  \qquad
     T_{F,1} (\vbr) =T_{F,2} (\vbr)  
                    =T_{F,3} (\vbr) 
                    =\frac{1}{5} (6 \pi^2)^{\frac{2}{3}} \hbar^2  
                     \rho_F (\vbr)^{\frac{5}{3}} .
\end{equation}

In order to simplify the analytic calculation, 
we introduce the dimensionless constants and variables:
\begin{eqnarray}
h = \omega_f \frac{M_B(M_B + M_F)}{M_F^2}\frac{a_{BF}}{a_{BB}},
                &\quad &
    \vx = \frac{4}{3 \pi}
                  \frac{M_F^4 \Omega_F^5 a_{BB}}{ \hbar M_B^{3} \Omega_B^{4}} 
                  ( \omega_T r_1, \omega_T r_2, \omega_L z ), 
\nonumber \\
 n_B = \frac{128}{9 \pi} \left(\frac{M_F}{M_B}\right)^8
                \omega_f^{10} a_{BB}^3  \rho_B, 
& \quad
     & n_F = \frac{32}{9 \pi}  \frac{M_F^9 (M_B + M_F)^3}{M_B^{12}}
                \omega_f^{12} a_{BB}^3 \rho_F, 
\label{scVal}
\end{eqnarray}
and the scaled dimensionless total energy of the ground state is written by
\begin{eqnarray}
     {\tilde E}_T^{(0)} &=& \frac{4^9}{3^7 \pi^{6} \hbar^4}
                            \frac{M_F^{28} \Omega_F^{35}}{M_B^{24}\Omega_B^{30}} 
                            a_{BB}^8  
                            E_T(\vlambda=0,{\dot \vlambda}=0)
\nonumber \\
                        &=& \int d^3{x} 
                            \Bigg\{ x^2 n_B  
                                   +\frac{1}{2} n_B^2 
                                   +\frac{3}{5} n_F^{\frac{5}{3}} 
                                   +x^2 n_F 
                                   +h n_B n_F \Bigg\} ,
\label{etotTFsg}
\end{eqnarray}
where $x^2 = |\vx|^2$. 

The TF equations become
\begin{equation}
     n_B +h n_F = e_B -x^2, \quad
     n_F^{\frac{2}{3}} +h n_B = e_F -x^2,
\label{TFeq}
\end{equation}
where $e_B$ and $e_F$ are the scaled boson and fermion chemical potentials.

\subsection{Collective Oscillations in Scaling Method}

We introducing the vector notation for the oscillation amplitudes:
\begin{equation}
\vlambda =\left(\begin{array}{c} \vlambda_B \\ \vlambda_F \end{array}\right), \quad
     \vlambda_B =\left(\begin{array}{c} \lambda_{BT} \\ \lambda_{BL} \end{array}\right), \quad
     \vlambda_F =\left(\begin{array}{c} \lambda_{FT} \\ \lambda_{FL} \end{array}\right), 
\label{veclambda}
\end{equation} 
where two-component matrices $\vlambda_B$ and $\vlambda_F$ are the bosonic and fermionic amplitudes.
Expanding the total energy (\ref{etotTFs})
to the order of $O(\lambda^2$), 
we obtain the oscillation energy in the harmonic approximation:
\begin{equation}
     \Delta {\tilde E}_T \equiv {\tilde E}_T - {\tilde E}^{(0)}_T 
                \approx \frac{1}{2} \tdvlamb B \dvlamb
                            +\frac{1}{2} \tvlamb C \vlambda.
\label{OsE}
\end{equation}
where $B$ and $C$ are $4 \times 4$ matrices defined as follows;
The matrices $B$ is 
\begin{equation}
     B = \frac{2}{3} 
         \left( \begin{array}{cc} 
                X_B  \tlB & 0  \\
                0 &  \frac{1}{\omega_f^2}  X_F \tlB  \\
                \end{array} \right) ,
\label{MTmx}
\end{equation}
where $\tlB$ is a diagonal matrix: 
$\tlB = {\rm diag}(2/\omega_T^2, 1/\omega_L^2)$
with $\omega_k \equiv \Omega_k / \Omega_B~~ (k=T,L)$.
And the matrix $C$ as 
\begin{eqnarray}
     C &=& \frac{2}{3} 
           \left( \begin{array}{cc} 
                  X_B \left(1-\frac{V_1}{10 X_B} \right) \tlC_B & V_1 \tlC_B  \\ 
                  V_1 \tlC_B                                    & X_F \tlC_F
                  \end{array} \right) .
\label{RFmx2}
\end{eqnarray}
The $2 \times 2$ submatrices $\tlC_B$ and $\tlC_F$ are 
\begin{equation}
     \tlC_B = \left( \begin{array}{cc} 
                     8  & 2 \\
                     2  & 3 
                     \end{array} \right), \qquad
     \tlC_F = \left( \begin{array}{cc} 
                     8 -\frac{4 V_1}{5 X_F} & -\frac{V_1}{5 X_F} -\frac{V_2}{X_F} \\
                    -\frac{V_1}{5 X_F} -\frac{V_2}{X_F} & 4 -\frac{3
		    V_1}{10 X_F} + \frac{V_2}{2 X_F} \\
                     \end{array} \right).
\label{RFmxEL}
\end{equation}
with
\begin{eqnarray}
    X_{B,F} &=& \int d^3{x}\, x^2 n_{B,F}(\vx), \\
     V_1 &=& h \int d^3{x}~ x^2~ \frac{\partial n_B}{\partial x} 
                    \frac{\partial n_F}{\partial x}, 
\label{VEL1}\\
     V_2 &=& h \int d^3{x} ~x ~ n_F \frac{\partial n_B}{\partial x}.
\label{VEL}
\end{eqnarray}
where $x^2 = |\vx|^2$.

From Eq.~(\ref{OsE}), 
the oscillation frequency $\omega$ 
and the corresponding amplitude $\vlambda$ are determined 
from the characteristic equation:
\begin{equation}
     \left( B \omega^2 - C  \right) \vlambda  = 0.
\label{eigEq}
\end{equation}
The above equation has four kinds of eigenvectors $\vlambda_i$ and the corresponding eigen-frequencies $\omega_i$.
which satisfies the orthogonal relation, 
$^t \vlambda_i B \vlambda_j \propto \delta_{ij}$.

For later convenience, 
we introduce some intrinsic oscillation frequencies defined 
from the diagonal components of the matrices $B$ and $C$:
\begin{equation}
     \omega_i = \sqrt{ \frac{C_{ii}}{B_{ii}} }, \quad
     i=1,2,3,4.
\label{intFr0}
\end{equation}
which correspond to the boson-transverse (BT),
the boson-longitudinal (BL), 
the fermion-transverse (FT), 
and the fermion-longitudinal (FL) modes.
It should be noted that these modes become eigenvalues 
of the characteristic equation (\ref{eigEq}) 
if the non-diagonal matrix elements are neglected (as it were, diabatic modes).
with $i=$BT (boson-transverse mode), BL (boson-longitudinal mode).
In substitution of the diagonal elements in  Eqs. (\ref{MTmx}) and
(\ref{RFmx2}) into Eq.~(\ref{intFr0}), simple calculation  gives
\begin{equation}
\begin{array}{lll}
     & \omega_{BT}^2 = 4 \omega_T^2 \left(1-\frac{v_B}{10} \right),
     & \omega_{BL}^2 = 3 \omega_L^2 \left(1-\frac{v_B}{10} \right),  \qquad\\
     & \omega_{FT}^2 = 4  \omega_T^2 \omega_f^2 \left(1-\frac{v_f}{10} \right),
     & \omega_{FL}^2 = \omega_L^2 \omega_f^2 
                      \left(4-\frac{3v_F}{10}+\frac{v_2}{2} \right) 
\end{array}
\label{intFr}
\end{equation}
with $v_B = V_1/X_B$, $v_F = V_1/X_F$, $v_2 = V_2/X_B$, 
and the corresponding amplitudes are
\begin{equation*}
     \vlambda_{BT} =\left( \begin{array}{c} \vu_T \\ 0 \end{array}\right), \quad
     \vlambda_{BL} =\left( \begin{array}{c} \vu_L \\ 0 \end{array}\right), \quad
     \vlambda_{FT} =\left( \begin{array}{c} 0 \\ \vu_T \end{array}\right), \quad
     \vlambda_{FL} =\left( \begin{array}{c} 0 \\ \vu_L
			     \end{array}\right) ,
\end{equation*}
where $\vu_T ={}^t(1,0)$ and $\vu_L ={}^t(0,1)$.

The other intrinsic modes are the bosonic and fermionic modes 
obtained by diagonalizing the bosonic and fermionic parts of the characteristic equation (\ref{eigEq}) \cite{MY12}:
\begin{equation}
     [\tlB \omega^2 - (1-v_B/10) \tlC_B] \vlambda_B = 0, \quad
     [\tlB \omega^2 - \omega_f^2\tlC_F] \vlambda_F = 0,
\label{intBFeq}
\end{equation}
which are the oscillations for the frozen boson or fermion distributions.

In the following subsections
we consider these modes in three extreme cases,
spherical, largely prolate, and oblate deformed traps, 
and clarify the aspect of the boson and fermion oscillation mixings, 
which was not clearly given in Ref.~\cite{MY12}.

\subsection{Collective Modes in Spherical Trap}

First we consider the collective modes of the BF mixture in the spherical trap
where $\omega_T = \omega_L = 1$.
To simplify the discussion, we take the same trap
frequencies of the boson and fermion, $\omega_f =1$. 
In this case, the oscillations are decoupled into 
the monopole an quadrupole modes, 
where $\vlambda_{B,F} \propto \vu_M \propto {}^t(1,1)$ and 
$\vlambda_{B,F} \propto \vu_Q \propto {}^t(1,-2)$, respectively.

The monopole oscillation has two modes: 
\begin{gather*}
     \omega_M^2 (\pm 1) =\frac{1}{2} 
                       \left( 9 - \frac{v_B}{2}
                             -\frac{v_F}{2} -\frac{v_2}{2} 
                             \pm \sqrt{D_M} \right), \\
     D_M = \left( 1 -\frac{v_B}{2} +\frac{v_F}{2}
                    +\frac{v_2}{2} \right)^2 + v_B v_F  .
\end{gather*}
The amplitudes corresponding to $\omega(\pm 1)$ are 
\begin{eqnarray*}
 \vlambda_M(+1) &=& \left[ \begin{array}{c}
    \left( 1 -\frac{v_B}{2}  +\frac{v_F}{2} +\frac{v_2}{2} 
                      +\sqrt{D_M} \right) \vu_M \\ 
                                v_F \vu_M
                                \end{array}\right], \\
 \vlambda_M(-1) &=& \left[ \begin{array}{c}
                              -v_B \vu_M \\
                              \left( 1 -\frac{v_B}{2} 
                                    +\frac{v_F}{2} +\frac{v_2}{2}
                                    + \sqrt{D_M} \right) \vu_M 
                              \end{array}\right].
\end{eqnarray*}

The quadrupole oscillation has also two modes:
\begin{gather*}
     \omega_Q^2 (\pm 1) = \frac{1}{2} 
                        \left( 2 -\frac{v_F}{5} 
                              -\frac{v_B}{5} + v_2 
                              \pm \sqrt{D_Q} \right), \\
     D_Q = \left( 2 + v_2 
                 -\frac{v_F}{5} +\frac{v_B}{5} \right)^2 
         + \frac{4}{25} v_B v_F,
\end{gather*}
and the corresponding amplitudes are
\begin{eqnarray*}
     \vlambda_Q(+1) &=& \left[ \begin{array}{c}
                               \frac{v_B}{5} \vu_Q \\
                               \left( 1+ \frac{v_B}{10} 
                                     -\frac{v_F}{10} +\frac{v_2}{2}
                                     +\frac{1}{2} \sqrt{D_Q} \right) \vu_Q 
                               \end{array}\right], \\
     \vlambda_Q(-1) &=& \left[ \begin{array}{c}
                                \left( 1 +\frac{v_B}{10} 
                                      -\frac{v_F}{10} +\frac{v_2}{2} 
                                      +\frac{1}{2} \sqrt{D_Q} \right) \vu_Q \\
                                -\frac{v_F}{5} \vu_Q 
                                \end{array}\right].
\end{eqnarray*}

Because of $\partial n_F/\partial x <0$ when $h <1$ and
 $\partial n_F/\partial x >0$ when $h >1$ in the boson-populated region, 
the signs of $v_B$ and $v_F$  become $v_{B,F} < 0$ 
when $h < 0$ or $h >1$, and  $v_{B,F} > 0$ 
when $0< h < 1$.

When  $h < 0$ or $h >1$, then,
 the $\vlambda_M(-1)$ and $\vlambda_Q(-1)$ are boson-fermion in-phase, 
and $\vlambda_M(+1)$ and $\vlambda_Q(+1)$ are out-of-phase,
while the relative phase is opposite when $0 <h < 1$.
Thus, in the case of the spherical trap, 
we find four collective oscillations:
the in-phase monopole (IM), the out-of-phase monocle (OM), 
the in-phase quadrupole (IQ), and the out-of-phase quadrupole (OQ) modes.

In the case of small BF coupling ($0 < h \ll 1$), 
the boson and fermion oscillation modes almost decouple.
In actual experiments and theoretical calculation discussed later, 
the boson number is set to be much larger than the fermion number 
($N_b \gg N_f$).
In the boson-dominant approximation, 
we can put $|v_B| \ll |v_F|$ because the integral $X_{B,F}$ satisfy the relation $X_B \gg X_F$, 
and the amplitudes becomes $\vlambda_M(+1) \propto {}^t(0,\vu_M)$ and
$\vlambda_Q(+1) \propto {}^t(0,\vu_Q)$ approximately.
So, in the boson-dominant approximation,
the OM and IQ modes become similar 
with the fermion monopole (FM) and fermion quadruple (FQ) oscillations.

When $-1 \ll h < 0$, the same argument is available though the relative phase
between the boson and fermion oscillations is opposite.

\subsection{Collective Modes in Largely Prolate Deformed Traps}

We discuss the collective oscillation modes in largely deformed traps
($\omega_T \gg \omega_L$) when $\omega_f = 1$.

Before discussing the collective oscillation in the full-calculation, 
we show the bosonic and fermionic intrinsic modes obtained by solving 
(\ref{intBFeq}). 

The boson oscillation has two kinds of oscillation:
the boson transverse breathing (BTB) and
the boson axial breathing (BAB) modes.
The frequencies are
\begin{equation}
     \omega_{BTB}^2 \approx 4 \left( 1 -\frac{v_B}{10} \right) \omega_T^2, \quad
     \omega_{BAB}^2 \approx \frac{5}{2} \left( 1 -\frac{v_B}{10} \right) \omega_L^2,
\label{frBTBBAB}
\end{equation}
and the corresponding amplitudes are
$\vlambda_B ~ \approx ~ \vu_T$ and
$\vlambda_B ~\approx~ \vu_{BA} \equiv {}^t(-1/4,~1)$ \cite{MY12}.

The fermionic oscillation also has two modes:
the fermion transverse breathing (FTB) 
and the fermion axial breathing (FAB) modes.
The frequencies are
\begin{equation}
     \omega_{FTB}^2 \approx  4 \left( 1 -\frac{v_F}{10} \right) \omega_T^2, \quad
     \omega_{FAB}^2 \approx \left( 4 + \frac{15 v_2 -12 v_F -4 v_2 v_F +v_F^2 }{40 - 4v_F} \right) \omega_L^2,
\label{frFTBFAB}
\end{equation}
and the corresponding amplitudes are
\begin{equation*}
    \vlambda_F \approx u_{T}, \quad
    \vlambda_F \approx u_{FA} 
               \equiv \left( \begin{array}{c} -\frac{v_2 +v_F/5}{4 -2v_F/5} \\ 1 \end{array} \right).
\end{equation*}

Now we turn to the collective oscillations obtained by solving (\ref{eigEq}).
In the full calculation, 
we find that the transverse and axial oscillation modes do not mix in the limit of $\omega_L/\omega_T \rightarrow 0$,
so that the mode-mixing of the BTB and FTB, and the BAB and FAB occurs,

The resultant transverse oscillations are 
the in-phase and out-of-phase transverse breathing (ITB and OTB) modes, 
the frequencies and the amplitudes of which are
\begin{equation*}
     \omega_{ITB}^2 \approx 4 \omega_T^2, \quad
     \omega_{OTB}^2 \approx \left[ 4 -\frac{2}{5} (v_B+v_F) \right] \omega_T^2,
\end{equation*}
and
\begin{equation*}
     \vlambda_{ITB} \approx \left( \begin{array}{c} \vu_T \\ \vu_T  \end{array} \right), \quad
     \vlambda_{OTB} \approx \left( \begin{array}{c} -\frac{X_B}{X_F} \vu_T \\ \vu_T \end{array} \right) 
              \approx \left( \begin{array}{c} 0 \\ \vu_T \end{array} \right).
\end{equation*}
The ITB mode has the same amplitudes of the boson and fermion oscillations, 
while the OTB mode is very similar to the FTB mode 
in the boson-dominant approximation ($X_B \ll X_F$). 

The axial breathing (AB) oscillations have also two modes; 
their frequencies are
\begin{gather*}
 \frac{\omega^2 (s)}{\omega_L^2} =\frac{13}{4} - \frac{v_B}{8} 
     +\frac{ 15v_2 - 12 v_F  - 4v_2 v_F + v_F^2 }{80(1 - \frac{v_F}{10})}
   + \frac{s}{2} \sqrt{D_{A}}, \\
     D_{A} =\left\{ \frac{3}{2} + \frac{v_B}{4} 
                   +\frac{ 15v_2 - 12v_F - 4v_2 v_F + v_F^2 }{40 \left(1 - \frac{v_F}{10} \right)} \right\}^2 
                   +\frac{v_B v_F}{4},
\end{gather*}
where $s=\pm 1$, and the corresponding amplitudes, $\vlambda_{AB} (s)$, are
\begin{eqnarray*}
  \vlambda_{AB} (+1) &\approx& \left[ \begin{array}{c} 
   v_{B} u_{BA} \\  
      \left(3 + \frac{v_B}{2} 
         +\frac{ 15v_2 - 12 v_F  - 4v_2 v_F + v_F^2 }{20 - 2 v_F} 
                      +2 \sqrt{D_A} \right) u_{FA} 
                                     \end{array} 
                              \right], \\
  \vlambda_{AB} (-1) &\approx& \left[ \begin{array}{c} 
               \left( 3 + \frac{v_B}{2} 
            +\frac{ 15v_2 - 12 v_F  - 4v_2 v_F + v_F^2 }{20 - 2 v_F} 
                           +2 \sqrt{D_A} \right) u_{BA} \\
                                     - v_F  u_{FA} 
                                       \end{array} 
                               \right].
\end{eqnarray*}
The states with $s=1$ and $s=-1$ are out-of-phase and in-phase,
respectively, when $h<0$ or $1<h$, 
and they are opposite when $0<h<1$.
In the boson-dominant approximation, in addition, 
the oscillation mode with $s=1$ becomes similar with the FAB oscillation 
because of $X_B \ll X_F$ and $|\nu_B| \ll |\nu_F|$.

\subsection{Collective Oscillation Modes in Largely Oblate Deformed Traps}

Next we consider the mixtures in largely oblate deformed traps ($\omega_L \gg \omega_T$) when $\omega_f = 1$.

In this limit, solving (\ref{intBFeq}),
we have two bosonic and two fermionic intrinsic modes.
Two bosonic oscillations are
the boson longitudinal breathing (BLB) and 
the boson sidewards breathing (BSB) modes:
\begin{equation*}
     \omega_{BLB}^2 \approx 3 \left( 1 -\frac{v_B}{10} \right) \omega_L^2, \quad
     \omega_{BSB}^2 \approx \frac{10}{3} \left( 1 -\frac{v_B}{10} \right) \omega_T^2.
\end{equation*}
The corresponding amplitudes are
$\vlambda_{BLB} \approx u_{L} \equiv {}^t(0, ~1)$ and  
$\vlambda_{BSB} \approx u_{BS} \equiv {}^t(1,~2/3)$. 
The BTB mode is a purely longitudinal oscillation, 
while the BSB mode includes the longitudinal and transverse oscillations 
with out-of-phases.

The fermion oscillation has also two modes:
the fermion longitudinal breathing (FLB) and 
the boson sidewards breathing (FSB) modes
with the frequencies
\begin{eqnarray*}
     \omega_{FLB}^2 &\approx& \left( 4 +\frac{v_2}{2} -\frac{3}{10} v_F \right) \omega_L^2, \\
     \omega_{FSB}^2 &\approx& 4 \left\{ 1 -\frac{v_F}{10}
                                  -\frac{\left( v_2 + \frac{v_F}{5} \right)^2}{32 + 4v_2 - \frac{12 v_F}{5} )} 
                              \right\} \omega_T^2,
\end{eqnarray*}
and the amplitudes:
\begin{equation*}
     \vlambda_{FLB} \approx \vu_L, \quad
     \vlambda_{FSB} \approx \vu_{FS} 
                \approx \left( \begin{array}{c} 
                                1 \\ 
                                \frac{40 v_2 + 8 v_F }{40 + 5 v_2 - 3 v_F}
                                \end{array} 
                         \right).
\end{equation*}

Now we show the collective oscillation in the case of largely oblate trap in full calculation,
where longitudinal and the sideward-oscillation modes do not mix 
in the limit of $\omega_T/\omega_L \rightarrow 0$.

The longitudinal oscillation has two modes:
the in-phase and out-of-phase longitudinal breathing (ILB and OLB) modes,
the frequencies of which are
\begin{gather*}
\frac{\omega^2(s)}{\omega_L^2}  =
\frac{7}{2}  -\frac{3v_B}{10} -\frac{3 v_F}{10} 
                        +\frac{v_2}{2}
                             + s \sqrt{D_{L}} , \\
     D_{L} =\left( 1 +\frac{v_2}{2} 
                     -\frac{3 v_F}{10} 
                     +\frac{3 v_B}{10} \right)^2 
           +\frac{36}{100}   v_B v_F,
\end{gather*}
where $s=\pm1$,
and the corresponding amplitudes, $\vlambda_{LB}(s)$, are
\begin{eqnarray*}
     \vlambda_{LB}(+1) &\approx& \left[ \begin{array}{c} 
                                     \frac{3}{5} v_B \vu_L \\
                                     \left( 1 +\frac{3v_B}{10} 
                                              +\frac{v_2}{2} 
                                              -\frac{3 v_F}{10}
                                              +\sqrt{D_{L}} 
                                     \right) \vu_L
                                     \end{array} 
                                     \right], \\
     \vlambda_{LB}(-1) &\approx& \left[ \begin{array}{c}
                                   \left( 1 +\frac{3v_B}{10} 
                                            +\frac{v_2}{2} 
                                            -\frac{3 v_F}{10}
                                            +\sqrt{D_L} 
                                    \right) \vu_L \\
                                    -\frac{3}{5} v_F \vu_L 
                                    \end{array} 
                            \right].
\end{eqnarray*}

The sidewards oscillation also has two modes:
the in-phase and out-of-phase sidewards breathing (ISB and OSB) modes,
The frequencies are given by
\begin{gather*}
\frac{ \omega^2 (s)}{\omega_T^2}
  =  \frac{11}{3}  -\frac{v_B}{6} 6  -\frac{v_F}{5}
          -\frac{\left( v_2 + \frac{v_F}{5} \right)^2}{16 + 2v_2 - \frac{6 v_F}{5} )} 
          \pm \frac{s}{2} \sqrt{D_{S}} , \\
     D_S =\left\{ \frac{2}{3} 
                 +\frac{v_B}{3} 
                 +\frac{2v_F}{5}
                 -\frac{\left( v_2 + \frac{v_F}{5} \right)^2}{8 +v_2 -\frac{3 v_F}{5} )} 
          \right\}^2
         +\frac{9 v_B v_F}{25} ,
\end{gather*}
where $s=\pm 1$,
and the corresponding eigenvectors, $\vlambda_{SB}(s)$, are
\begin{eqnarray*}
     \lambda_{SB}(+1) &\propto& \left[ \begin{array}{c}  
                                  \frac{9}{10} v_B \vu_{BS} \\
                                  \left\{ 1 +\frac{v_B}{2} 
                                            -\frac{3v_F}{5}
                                            +\frac{3\left( v_2 +\frac{v_F}{5} \right)^2}{16 (1 +\frac{v_2}{8} -\frac{3 v_F}{40})} 
                                            +\frac{3}{2}\sqrt{D_{S}} 
                                  \right\} \vu_{FS}
                                  \end{array} \right], \\
    \lambda_{SB}(-1) & \propto & \left[ \begin{array}{c}  
                                   \left\{ 1 +\frac{v_B}{2} 
                                             -\frac{3v_F}{5}
                                             +\frac{3\left( v_2 +\frac{v_F}{5} \right)^2}{16 (1 +\frac{v_2}{8} -\frac{3 v_F}{40})} 
                                             +\frac{3}{2}\sqrt{D_{S}} 
                                   \right\} \vu_{BS} \\
                                   -\frac{9}{10} v_F \vu_{FS} 
                                   \end{array}
                           \right] .
\end{eqnarray*}

In this limit
the states with $s=1$ and $s=-1$ are out-of-phase and in-phase,
respectively, when $h<0$ or $1<h$, 
and they are opposite when $0<h<1$.
In the boson-dominant approximation
the oscillation mode with $s=1$ becomes similar with the FAB oscillation

\bigskip

In all the limits, thus, the oscillation states are classified into 
two kinds of modes in the boson-dominant approximation. 
One is the boson and fermion co-moving, and the other is
a fermion moving almost alone.
When a large part of the condensed bosons occupy a single-particle state, 
the boson oscillation has the one-sided effect on the fermion oscillation. 
It is a characteristic properties of the collective oscillation 
of the BF mixture 
and has been seen also in the time-dependent approach \cite{tomoBF,QPBF}.  

\newpage

\section{Numerical Results and Discussions for Oscillation modes}

In this section, we show the full calculation of the frequencies and 
amplitudes of the collective oscillation in the BF mixture obtained numerically.
In actual calculations, we take the Yb-Yb system, 
where the numbers of the bosons and the fermions 
are $N_b = 30000$ and $N_f = 1000$, respectively.
The parameters of the trapping potential are
$\Omega_B = 2 \pi \times (300^2 \times 50)^{\frac{1}{3}}$Hz and $\omega_f = 1$;
these values are almost similar with those in the experiments by Kyoto group \cite{Fukuhara-M} 
(the axial symmetry is a little broken in the actual experiment).
The mass differences among Yb-isotopes can be safely neglected, 
so that we use the same mass and the same trap frequency ($\omega_f =1$) 
for all Yb-isotopes  in the calculation.

\begin{figure}[h]
\begin{center}
{\includegraphics[scale=0.65,angle=270]{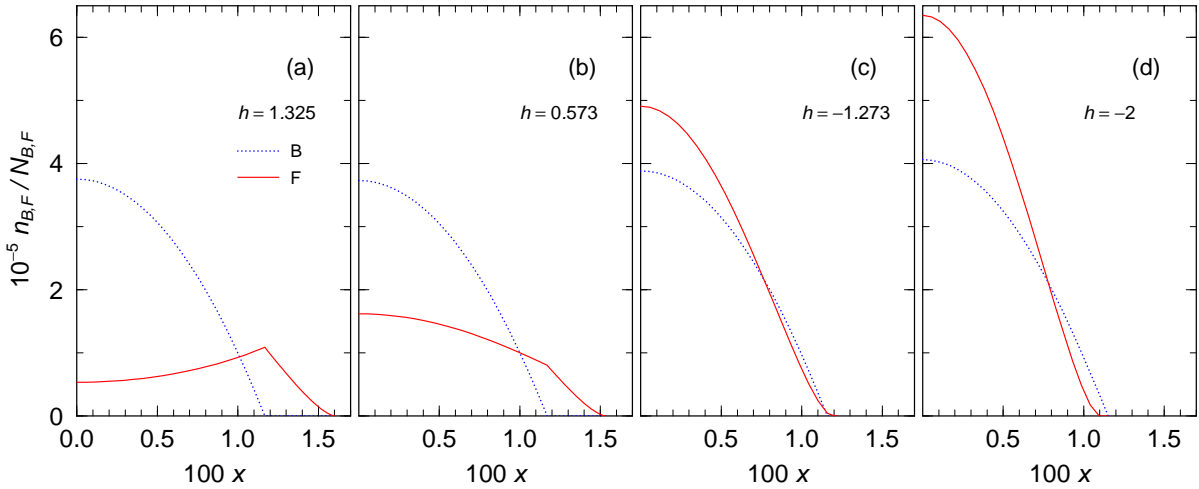}}
\caption{\small (Color online)
The ground-state density distributions of the BF mixtures
in the TF approximation
for $h = 1.325$ (a),  $0.573$ (b),  $-1.273$ (c) and 
$-2.5$ (d).
The dotted and solid lines are for the boson and fermion 
distributions.
The numbers of the bosons and the fermions 
are $N_b = 30000$ and $N_f = 1000$, respectively.
The parameters of the trapping potential are
$\Omega_B = \Omega_F = 2 \pi \times (300^2 \times 50)^{\frac{1}{3}}$Hz.}
\label{grdRH}
\end{center}
\end{figure}

In Fig.~\ref{grdRH}, 
we show the scaled ground-state density distributions 
of boson (dotted line) and fermion (solid lines) 
for $h = 1.325$ (a), $0.573$ (b), $-1.273$ (c) and $-2.0$ (d).

The boson densities are found to be center-peaked in all cases;
in contrast, the fermion densities are surface-peaked  
in $h > 1$ (Fig.~\ref{grdRH}a),
and center-peaked in $h < 1$. 
As the BF interactions becomes attractively stronger ($h < 0$),
the fermions distribute more largely in the boson-distributed regions, 
and the boson density increases at the center \cite{tbDPL,QPBF,MY12}.

For the boson number of $N_b=30000$,
the fermi gas is almost covered with the bose gas 
when $h<-1$.
When $N_b = 10000$ \cite{MY12}, 
the bose gas partially distributes outside the boson region.
The mixtures with large overlap region is expected 
to show significant behaviors of the collective oscillations more clearly, 
which is discuss later.

\begin{figure}[htb]
\begin{center}
{\includegraphics[scale=0.62,angle=270]{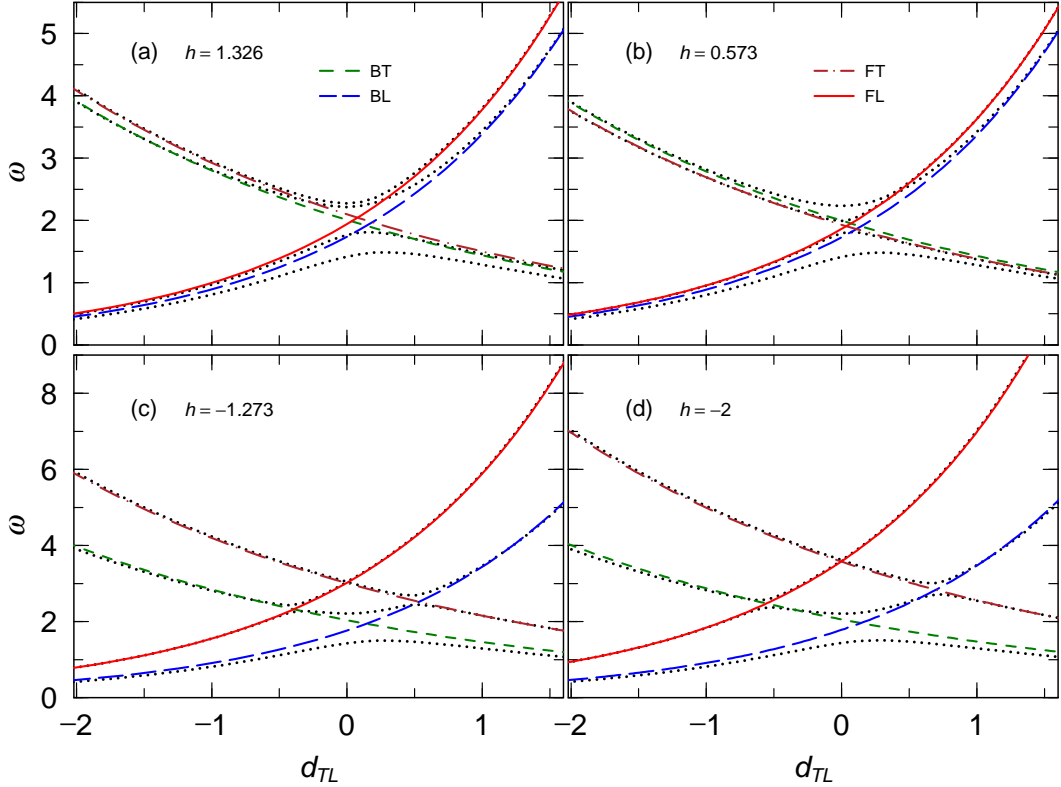}}
\caption{\small (Color online)
The frequencies of the collective oscillations versus the deformation parameters 
$d_{TL}$, which is defined in (\ref{dTL}),
when $h=1.326$ (a) 
and $h= 0.573$ (b), $h=-1.273$ (c) and $h=-2$ (d).
The dotted lines are for the frequencies in the full calculation 
in Eq.(\ref{eigEq}).
The dashed, long-dashed, dot-dashed and solid lines 
denote the frequencies of the boson-transverse, boson-longitudinal,
fermion-transverse and fermion longitudinal oscillations in (\ref{intFr}).}
\label{frYb}
\end{center}
\end{figure}

In order to characterize the deformation of the system, 
we introduce the deformation parameter:
\begin{equation}
     d_{TL} = \log\frac{\omega_L}{\omega_T} ,
\label{dTL}
\end{equation}
which is $d_{TL} >0$ for prolate shapes, $d_{TL} =0$ for spherical shapes, 
and $d_{TL} <0$ for oblate shapes.

In Fig.~\ref{frYb}, 
we show the $d_{TL}$-dependences of the frequencies of the breathing oscillations 
(dotted lines) solved in Eq.~(\ref{eigEq}) for $h=1.3255, 0.573, -1.273, -2$.
We call the corresponding states as the state-1,2,3,4 in decreasing
order of magnitudes.
The intrinsic frequencies obtained in (\ref{intFr}) are also plotted 
in Fig.~\ref{frYb}, 
and, of course, they have crossing points.
In Fig.~\ref{frYb}, the level repulsion is not clear 
between the states-1 and -2 for $h <0$ in the full calculations.
However, the existence of the very small level gap is confirmed 
in more precise calculation, so that the frequencies represented with
the dotted line in Fig.~\ref{frYb} show no crossings

We find that the intrinsic oscillation frequencies agrees 
with the full calculations  
except in the nearly-spherical region ($d_{TL} \approx 0$)  and at
the crossing points of the intrinsic frequencies, 
where the critical changes of the collective frequency behaviors (dotted
lines) occur:
it is just level crossing phenomena.

\begin{figure}[hbt]
\begin{center}
{\includegraphics[angle=270,scale=0.5]{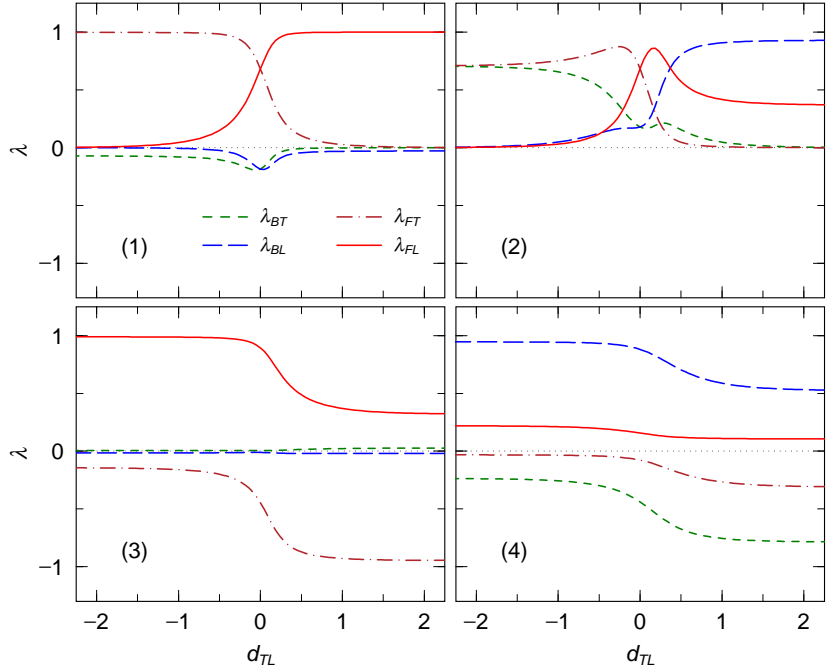}}
\caption{\small (Color online)
Components of the amplitudes
for the state-1 (1), state-2 (2), state-3 (3) and state-4 (4)
with $h = 1.326$.
The dashed and long dashed lines represent the transverse and 
longitudinal components of the boson oscillations, respectively,
and the dash-dotted and solid lines denote the transverse and 
longitudinal components of the fermion oscillations, respectively. 
 }
\label{svec1}
\end{center}
\end{figure}

\begin{figure}
\begin{center}
{\includegraphics[angle=270,scale=0.5]{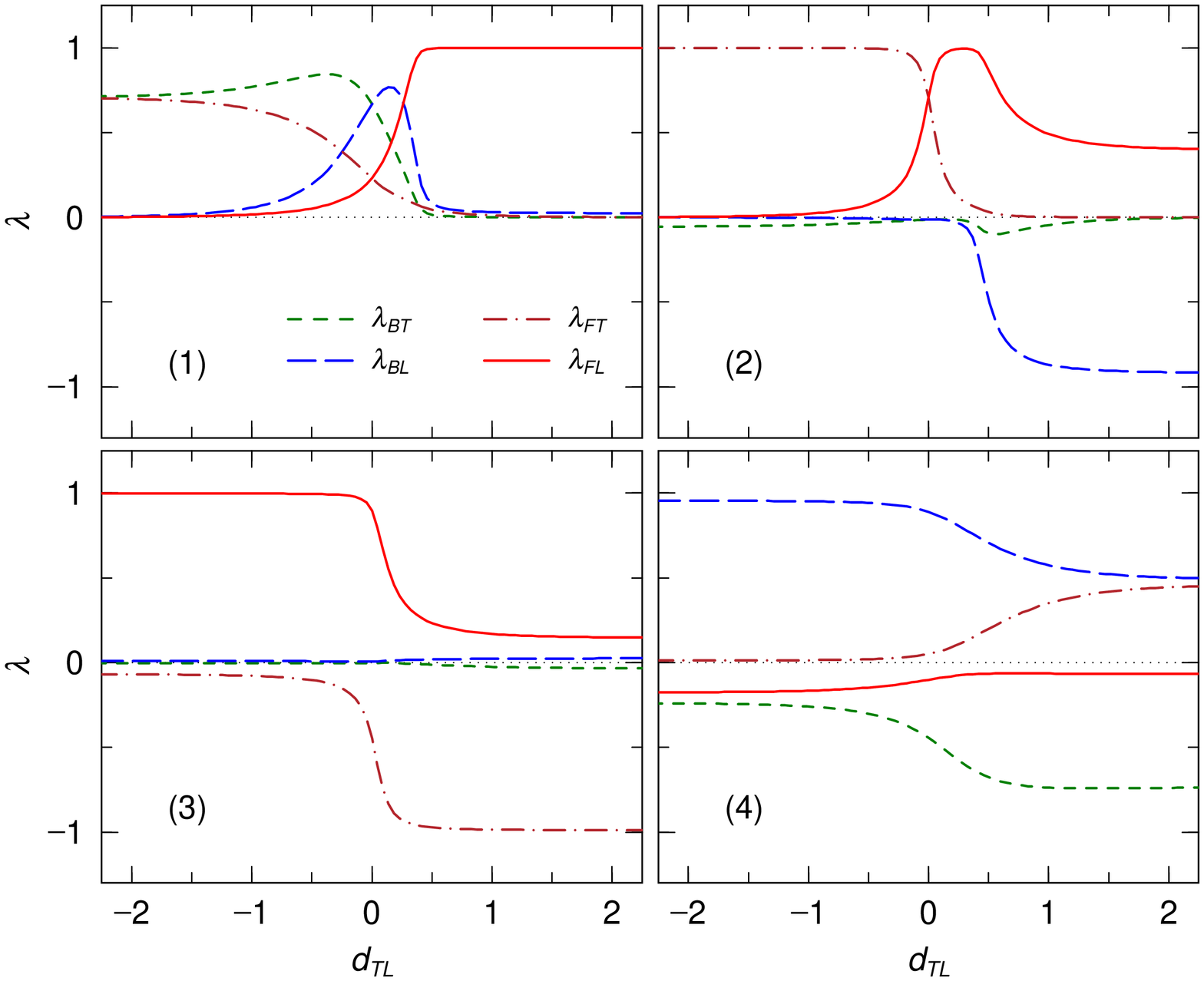}}
\caption{\small (Color online)
Components of the eigen-vectors
of state-1 (1), state-2 (2), state-3 (4) and state-4 (4)
with  $h=0.573$.
The meanings of the lines are the same as those in Fig.~\ref{svec1}.
 }
\label{svec2}
\end{center}
\end{figure}

\begin{figure}
\begin{center}
{\includegraphics[angle=270,scale=0.5]{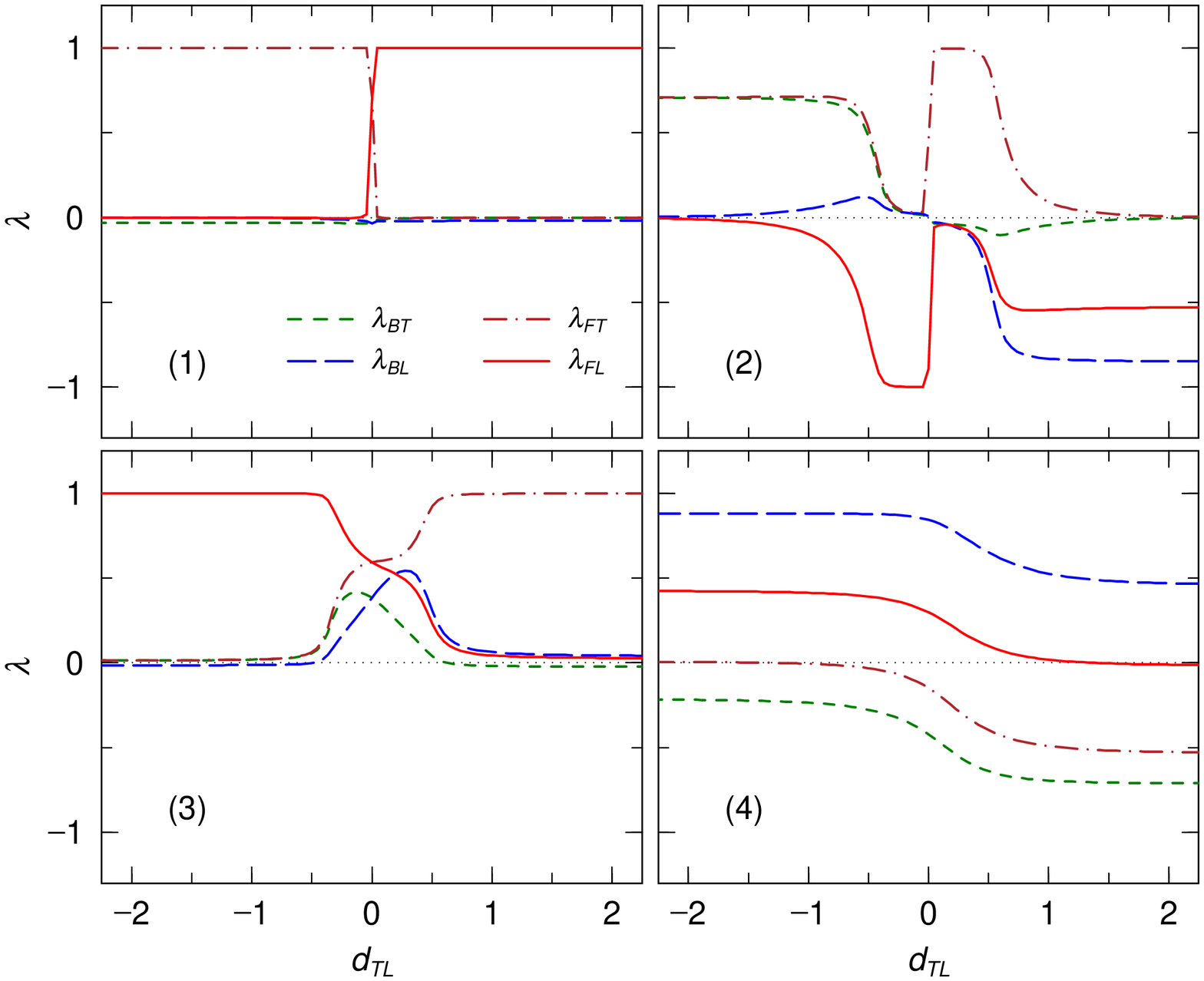}}
\caption{\small (Color online)
Components of the eigen-vectors
of state-1 (1), state-2 (2), state-3 (3) and state-4 (4)
with  $h = -1.273$.
The meanings of the lines are the same as those in Fig.~\ref{svec1}.
 }
\label{svec3}
\end{center}
\end{figure}

\begin{figure}[htb]
\begin{center}
{\includegraphics[angle=270,scale=0.5]{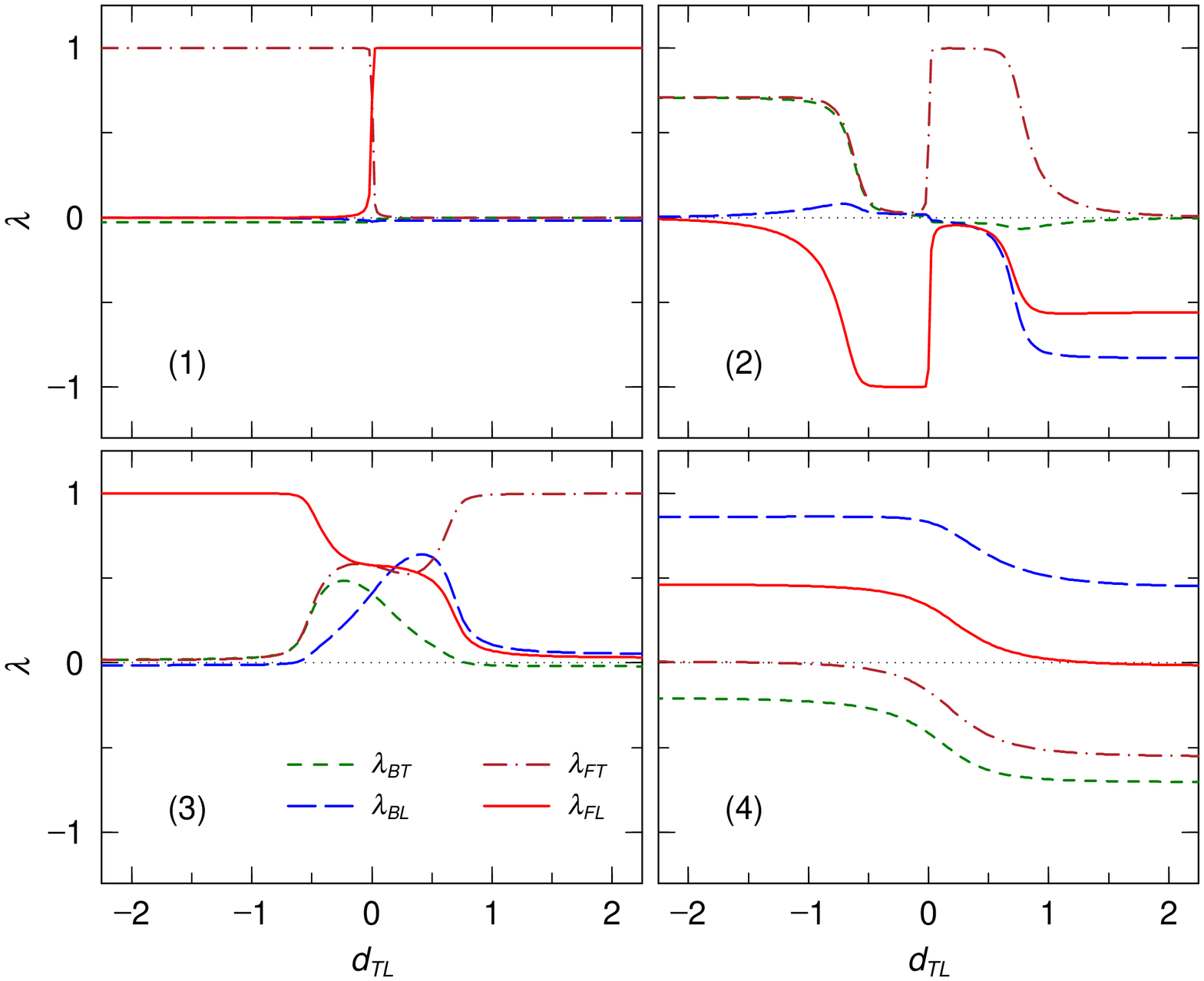}}
\caption{\small (Color online)
Components of the eigen-vectors
of state-1 (a), state-2 (b), state-3 (c) and state-4 (d)
with  $h = -2$.
The meanings of the lines are the same as those in Fig.~\ref{svec1}.
 }
\label{svec4}
\end{center}
\end{figure}

When $h=1.326$  (Fig.~\ref{frYb}a), 
the frequencies of the state-1 and -2 are very close for all deformation regions, 
and those of the state-3 and -4 are also close except the spherical region 
($d_{TL} \approx 0$).

In order to understand the characteristics of these states, 
we show the strengths of the intrinsic-mode components included 
in the amplitudes  for the state-1 $\sim$ -4 (the panels 1 $\sim$ 4)
for $h = 1.326$ (Fig.~\ref{svec1}), $0.573$ (Fig.~\ref{svec2}),  
$-1.273$ (Fig.~\ref{svec3}),  $-2$ (Fig.~\ref{svec4}).
The dashed and long-dashed lines represent the transverse ($\lambda_{BT}$) 
and longitudinal components ($\lambda_{BL}$) of the boson oscillations, 
respectively, and the dash-dotted and solid-lines denote 
the transverse ($\lambda_{FT}$) and longitudinal components ($\lambda_{FL}$) 
of the fermion oscillations, respectively. 

From these figures, 
we can understand the character changes of the states 
in reference with the modes obtained in the previous sections.

As an example, we consider the character changes of the state-1 in $h=1.326$.
From the panel (1) of Fig.~\ref{svec1}, 
we find that the $\lambda_{FT}$ component is dominant 
in the largely prolate-deformed region ($d_{TL} \lesssim -1$), 
so that the state-1 is mainly in the FTB (OTB) mode.
In the spherical region ($d_{TL} \approx 0$), 
we find the $\lambda_{FT} \approx \lambda_{FL}$; 
the state is in the FM (fermion monopole) mode.
As explained in the previous section, 
it is really the OM (boson-fermion out-of-phase monopole) mode 
accompanying the small boson amplitudes; 
we can find it in the small dips in the $\lambda_{BT}$ and $\lambda_{BL}$ lines 
in Fig.~\ref{svec1}(1).
We can find easily that the state-1 is in the FLB mode 
in the oblate region ($d_{TL} \gg 1$).

We analyze  the character changes of the collective oscillations,
in the same way, and show the results  in Table.~\ref{CvD}.

\begin{table}[ht]
\begin{tabular}{|c|c|ccccc|}
\hline
~$h$~ & ~~~state~~~ & prolate &&spherical&&oblate\\
\hline
\multirow{4}{*}{~~$1.325$~~} & ~state-1~ 
&~~ FTB (OTB)~~ &  & OM &  & ~~FLB (OLB)  ~~ \\
&state-2 & ITB &  &~ IM ~ & & ~~ILB ~~ \\
&state-3 & FAB (OAB) &  & FQ(OQ) &  & ~~FSB (OSB) ~~ \\
&state-4 & IAB &  & IQ & & ~~ISB ~~ \\
\hline
\multirow{4}{*}{~~$0.573$~~} & ~state-1~ 
&~~ITB~~ &  & IM &  & ~~FLB (ILB)  ~~ \\
&state-2 & FTB (OTB) &  &~ FM (OM) ~ & & ~~OLB ~~ \\
&state-3 & FAB (IAB) &  & FQ (IQ) &  & ~~FSB (ISB) ~~ \\
&state-4 & OAB &  & OQ & & ~~OSB ~~ \\
\hline
\multirow{2}{*}{~~$-1.273$~~} & ~state-1~ 
&~~FTB (OTB)~~ &  & FM (OM) &  & ~~FLB(OLB) ~~ \\
&state-2 & ITB  & FAB~~ &~ FQ (OQ) ~~ & ~~ FTB & ~~ILB ~~ \\
\multirow{2}{*}{~~$-2$~~}& state-3 & FAB (OAB) &  & IM &  & ~~FSB (OSB) ~~ \\
&state-4 & IAB &  & IQ & & ~~ISB ~~ \\
\hline
\end{tabular}
\caption{\small Oscillation modes versus Deformation}
\label{CvD}
\end{table}

The deformation regions of the system are classified into
the prolate-deformed, semi-spherical and oblate-deformed ones.
Furthermore, the oscillation behaviors and the order of their energy levels
are also different in three regions of the BF interaction: 
$h>1$, $1 > h>0$ and $0 > h$. 

As discussed in the previous section, 
the relative phase between the boson and fermion oscillations
in the co-moving oscillation modes when $0 < h < 1$
show different behavior from that in other cases 
except the ITB and OTB modes in the largely prolate deformation.

When the BF interaction is repulsive, $h>0$, the mode changes are
simple, and the in-phase and out-of phase modes do not appear in the states.

When the BF interaction is attractive, $h < 0$, 
the mode changes become complicated, especially in the
semi-spherical deformation region.
When the deformation of the system becomes large (in either prolate or oblate shapes),
the oscillation behaviors are simple; 
the four states becomes the intrinsic modes 
which are defined in the previous section.

As the deformation approaches to the spherical one, 
the oscillation modes show drastic changes 
between the fermion-dominant and the Bf co-moving modes, 
and between the in-phase and out-of-phase modes. 
The mixing of the oscillation modes varies complicatedly
in the transcendental regions around the spherical shape.

In addition, we should note that,
in the completely spherical system,
the oscillations are decoupled into the monopole and
quadrupole modes, 
but this decoupled behavior appears only in the very narrow region
around $d_{TL} \approx 0$. 

Thus, the oscillation modes in each state have the characters varying with the deformation of the trap.
We should comment that more drastic mode changes appear for the variation of the
BF-coupling $h$ is varied  \cite{tomoBF,QPBF,MY12} 

The ranges of the transcendental regions are approximately estimated 
as the region between the two crossing points at the deformation points
$d_1$ and $d_2$ in Fig.~\ref{frYb}; the crossing of  BLB and FTB
corresponds to $d_1$, and that of  BTB and FLB corresponds to $d_2$.

The points $d_1$ and $d_2$ are calculated from the crossing conditions 
$\omega_{BT} =\omega_{FL}$ and $\omega_{BL} =\omega_{FT}$, respectively.
Using Eqs. (\ref{intFr}), we obtain
\begin{eqnarray}
     d_1 &=&\frac{1}{2} 
            \log\left[
              \frac{4 \left(1 -\frac{v_F}{10}\right)}{
                    3 \left(1 -\frac{v_B}{10}\right) } 
               \right], \label{D1}\\
     d_2 &=&\frac{1}{2} 
            \log\left[
               \frac{4 \left(1 -\frac{v_F}{10} \right)}{
                     \left(4 -\frac{3v_F}{10} +\frac{v_2}{2}\right) }
              \right].
\label{D2}
\end{eqnarray}

\begin{figure}[htb]
\begin{center}
{\includegraphics[scale=0.5,angle=270]{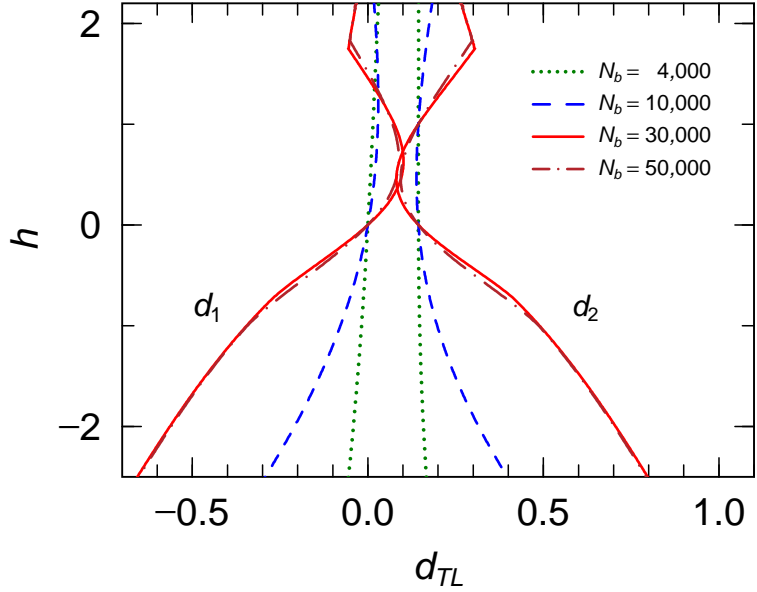}}
\caption{\small (Color online)
The level-crossing point of the deformation parameter
between the BLB and FTB modes, $d_1$, (a) and 
that between the BTB and FLB modes, $d_2$ (b), when $N_f = 1000$.
The dotted and dashed, solid and dotted-dashed lines represent
the results when $N_b = 4000$, $10,000$, $30,000$ and $50,000$, respectively.
 }
\label{defLC}
\end{center}
\end{figure}

In Fig.~\ref{defLC}, 
we show the level-crossing points $d_1$  and $d_2$ (a and b) 
when $N_b = 4000$ (dotted line), 
$N_b=10000$ (dashed line), $N_b=30000$ (solid line) and 
$N_b=50000$ (dot-dashed line), respectively.
When $N_b \gtrsim 30000$, 
the $d_1$ and $d_2$ are found to be are almost independent of $N_b$.
In addition, the order of $d_1$ and  $d_2$ are exchanged in 
$0.37 \lesssim h \lesssim 0.73$ when $N_b \lesssim 30000$.

Let's give the qualitative explanations of the $N_b$-independence 
of the critical points in the boson-dominate approximation ($N_b \gg N_f$).
When $N_b \gg N_f$, 
the boson density distribution becomes robust 
and is not affected so much by the fermion density, 
so that it becomes 
$n_B(x) \approx e_B - x^2$.
In the case of $h < 0$, 
the fermions populate inside the boson region, 
and the TF equation (\ref{TFeq}) gives
the fermion density distribution:
\begin{equation*}
     n_F \approx \left\{ (e_F - h e_B) -(1-h)x^2 \right\}^\frac{3}{2},
\end{equation*}
which is determined by $e_F -h e_B$, 
whose value is also decided by the fermion number.
In $ 0 < h \ll 1$ (the very weak BF interaction), 
the large part of fermions populate in the boson region, 
and the argument for $h <0$ is also available.

Next, when $h \approx 1$, 
$\partial n_F/\partial x \approx 0$ and $\partial n_B/\partial x \approx -2x$ 
inside the boson region, so that, using (\ref{VEL1}),
we obtain $V_1  \approx 0$, 
which leads to $\nu_B =V_1/X_B \approx 0$ and $\nu_F =V_1/X_F \approx 0$.
Substituting these values into (\ref{D1}) and (\ref{D2}), 
we obtain $d_2 \approx \log (4/3) / 2 \approx 0.14$.
Because   inside the boson region 
in addition,  $V_2 \rightarrow 0$ when $N_f/N_b \rightarrow \infty$,
and   $V_2 \rightarrow -2h X_F$ when $N_f/N_b \rightarrow 0$, namely
$\nu_2  \approx 0$ when $N_b \ll N_f$ and
$\nu_2  \approx -2$ when $N_b \gg N_f$.
Thus, we obtain $d_1 \approx 0$ when $N_b \ll N_f$, and 
$d_2 \approx \log (4/3) / 2 \approx d_1$ when $N_b \gg N_f$.

Finally, in $h > 1$, 
the fermion density distribution has the surface-peaked peak, 
and a large part of fermions populate outside the boson region.
The oscillation behavior is also determined by the density distribution 
around the boson surface, 
which does not so strongly depend on the boson numbers.
Thus, the level-crossing points are almost independent 
of $N_b$ when $N_b \gg N_f$. 

We should note that if $N_b \approx N_f$ or $N_b \ll N_f$, the 
fermions distribute largely outside of the boson region, and the overlap
region of the bosons and fermions is very small.
Then, the bose gas and fermi gas oscillate almost independently \cite{sogo}
and their oscillation behaviors are not so interesting.

The critical phenomena originated in the level crossing phenomena 
are seen only for the large values of$d_{1,2}$.
Hence we should focus this study on the system 
with $N_b \gtrsim 30 N_f$ and $h < -1$, 
where the fermions distribute
inside the boson region. 

In this system, 
the boson motions are rarely affected by the fermions, 
lead to the forced oscillations of fermions,
which are the boson-fermion co-moving oscillations. 
On the other hand, 
the fermion oscillation does not largely affect the boson oscillation, 
and then there are other kinds of the intrinsic modes
where the fermions oscillate dominantly
with very weakly oscillations of bosons.
In this oscillation, 
the fermion potential is equivalent to the harmonic-oscillator potential 
with the shape of $h e_B +(1-h) x^2$.

When the deformation parameters is in the regions of $d_1 < d_{TL}$ and 
$d_{TL} < d_2$,
the oscillation behavior becomes the same as that in the extremely deformed limit. 
For $d_{TL} < d_{TL} < d_1$, 
the mixings of the oscillation modes occur and 
cause the variations of their characters.
Especially, 
in the state-2 and the state-3,
the oscillation behavior changes critically 
around $d_{TL} \approx d_1$ and $d_{TL} \approx d_2$. 

\section{Comparisons with the Dynamical Approach}

\begin{figure}[htb]
\vspace*{-0.5cm}
\begin{center}
{\includegraphics[scale=0.5,angle=270]{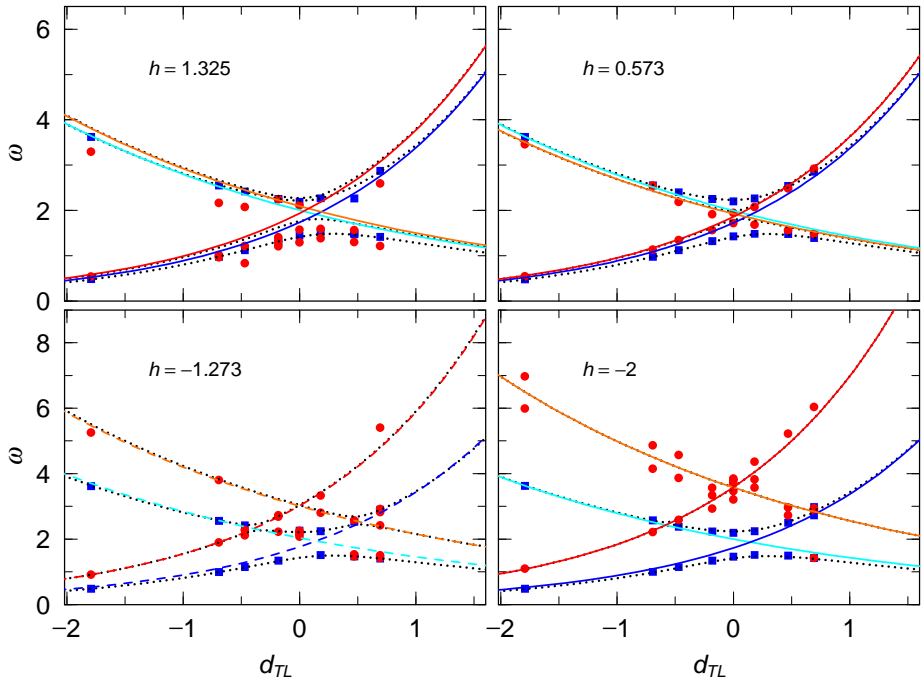}}
\caption{\small (Color online)
The frequencies of the collective oscillations in the BF mixtures 
obtained by the TDGP+Vlasov approach (solid squares and open circles).
The lines are the collective and the intrinsic frequencies obtained in the scaling method, 
which are the same ones in Fig.~\ref{frYb}.
}
\label{ScFrSm}
\end{center}
\end{figure}

In Fig.~\ref{ScFrSm}, we show the frequencies of the collective oscillation in the BF mixture 
calculated in the dynamical approach based on the TDGP+Vlasov equation; 
the derivations and the details of this method was given in our previous paper\cite{tomoBF,tbDPL,QPBF,MY12}.  
We find that these two methods give the consistent results except 
in the case of the strong boson-fermion attractive interactions.

We should comment that the dynamical calculations give consistent results with the RPA 
and its approximation like the scaling method 
in early stage of time development.
But it show discrepancies in later stage, 
and, as the boson-fermion interaction becomes stronger, 
the discrepancies appear earlier in the time-development.
Thus, the BF mixtures show new dynamical properties different from the other kinds of finite many-body system 
such as atomic nuclei.

\section{Summary}

In this paper, 
we have investigated the deformation dependence of the breathing oscillations 
in the boson-fermion mixture.
The deformation regions are classified into
the prolate, semi-spherical and oblate ones.
In each region, 
the four kinds of oscillations with different mixing characters, 
the BLB, BTB, FLB and FTB modes, appear;
around the borders of these regions, 
the level-crossing phenomena between the BTB and FLB and between the BTB and FLB modes occurs.

This classification based on the deformation of the system should be useful for experimentalists 
in the observation of these breathing
oscillations and/or related phenomena.
Furthermore, the oscillation behaviors become complicated in the
region between the largely deformed and spherical shapes, 
and the drastic changes occur at the level crossing points;
the resonant oscillations should appear at the level crossing points.

These variations of the oscillational behaviors are clearly seen 
in the case of the strong boson-fermion attractive interaction 
($h \lesssim -1$), and the mixtures of the large boson number ($N_b \gg N_f$),
where the boson-dominant approximation is available.

When $N_b \gg N_f$,
the oscillation states are mainly classified into two kinds:
boson-fermion co-moving modes and fermion-dominant oscillation modes.
The co-moving oscillation modes are actually the boson dominant oscillation
with the forced-oscillation of fermions; the
frequencies of the boson and fermion oscillations are the same.

In the time-dependent simulations \cite{tbDPL,QPBF,MY12},
the fermionic intrinsic oscillations show rapid damping, and,
particularly when $h \gtrsim 1$, their amplitudes are negligibly small.  
The strength of the fermion intrinsic oscillation is separated into 
the two oscillation modes corresponding to fermion motions inside and 
outside of  the boson populate region; \cite{tbDPL,QPBF,MY12}: 
these behaviors cannot be described with the scaling method.
When the BF-interaction is repulsive, large part of fermions populate
outside of the boson region, and, when $h>1$, the fermions outside the
boson region dominate.
Thus, the fermion dominant oscillation do not exist in long period, and
we cannot easily observe these  modes in actual experiments. 

On the other hand, 
the boson-fermion co-moving oscillations can be easily 
observed in experiments.
Indeed, the BLB oscillations have been found 
in the actual experiments \cite{Fukuhara3}, 
which showed the 1.2\% increase of its frequency 
when the BF coupling is varied from 
$h_{BF}/g_{BB} = 0$ to  $h_{BF}/g_{BB} = 1.325$.
In this experiment, 
because of the strongly-repulsive BF interaction, 
the fermion contribution should be small.
In the case of the strongly-attractive BF interaction ($h < -1$),
however, we can observe larger effects of the deformation in the oscillation behaviors,
and find that the clear differences between the spherical and non-spherical oscillations.  
The BF mixtures of the strongly-attractive BF interactions also
show the oscillation behaviors caused by the fermion-gas expansion \cite{monoEX,QPBF,MY12}.
This behavior cannot be included in the scaling method.

In order to obtain precise results of the oscillation behaviors of the BF mixtures, 
especially around the level-crossing points, 
the analysis based on the calculations of the time-dependent framework with the TDGP+Vlasov approach. 
The details of the deformation dependence of the time-development behaviors of the BF mixtures 
should be given in the other paper.

\bigskip
{\bf Acknowledgment}

This work is supported in part by the Japanese Grand-in-Aid for
Scientific Research Fund of the Ministry of Education, Science, Sports
and Culture (21540212 and 22540414) and 
Nihon University College of Bioresource Sciences Research Grant for 2013.

\end{document}